\documentclass[preprint,showkeys,showpacs,preprintnumbers,amsmath,pre,amssymb,tightenlines]{revtex4}
\usepackage{amsfonts}
\usepackage{graphicx}
\usepackage{natbib}
\usepackage{dcolumn}% Align table columns on decimal point
\usepackage{bm}% bold math
\usepackage{rotating}

\begin{document}

\preprint{}

\title{Clustering in Complex Directed Networks}

\author{Giorgio Fagiolo}
\email{giorgio.fagiolo@sssup.it} \affiliation{Sant'Anna School of
Advanced Studies, Laboratory of Economics and Management, Piazza
Martiri della Libert\`{a} 33, I-56127 Pisa, Italy.}

\date{December 2006}

\begin{abstract}
\noindent Many empirical networks display an inherent tendency to
cluster, i.e. to form circles of connected nodes. This feature is
typically measured by the clustering coefficient (CC). The CC,
originally introduced for binary, undirected graphs, has been
recently generalized to weighted, undirected networks. Here we
extend the CC to the case of (binary and weighted)
\textit{directed} networks and we compute its expected value for
random graphs. We distinguish between CCs that count all directed
triangles in the graph (independently of the direction of their
edges) and CCs that only consider particular types of directed
triangles (e.g., cycles). The main concepts are illustrated by
employing empirical data on world-trade flows.
\end{abstract}

\keywords{Clustering, Clustering Coefficient, Complex Networks,
Directed Graphs, Weighted Networks.}

\pacs{89.75.-k, 89.65.Gh, 87.23.Ge, 05.70.Ln, 05.40.-a}

\maketitle

Networked structures emerge almost ubiquitously in complex systems.
Examples include the Internet and the WWW, airline connections,
scientific collaborations and citations, trade and labor-market
contacts, friendship and other social relationships, business
relations and R\&S partnerships, cellular, ecological and neural
networks \cite{AlbertBarabasi2002,Newman2003,DoroMendes2003}.

The majority of such ``real-world'' networks have been shown to
display structural properties that are neither those of a random
graph \cite{Bollo1985}, nor those of regular lattices. For example,
many empirically-observed networks are small-worlds
\cite{Kochen1989,Watts1999}. These networks are simultaneously
characterized by two features \cite{Amaral2000}. First, as happens
for random graphs, their diameter \footnote{As computed by the
average shortest distance between any two nodes \cite{VanLee1999}.}
increases only logarithmically with the number of nodes. This means
that, even if the network is very large, any two seemingly unrelated
nodes can reach each other in a few steps. Second, as happens in
lattices, small-world networks are highly clustered, i.e. any two
neighbors of a given node have a probability of being themselves
neighbors which is much larger than in random graphs.

Network clustering is a well-known concept in sociology, where
notions such as ``cliques'' and ``transitive triads'' have been
widely employed \cite{WassermanFaust1994,Scott2000}. For example,
friendship networks are typically highly clustered (i.e. they
display high cliquishness) because any two friends of a person are
very likely to be friends.

The tendency of a network to form tightly connected neighborhoods
(more than in the random uncorrelated case) can be measured by the
clustering coefficient (CC), see \cite{WattsStrogatz1998} and
\cite{Szabo2004}. The idea is very simple. Consider a binary,
undirected network (BUN) described by the graph $G=(N,A)$, where $N$
is the number of the nodes and $A=\{a_{ij}\}$ is the $N\times N$
adjacency matrix, whose generic element $a_{ij}=1$ if and only if
there is an edge connecting nodes $i$ and $j$ (i.e. if they are
neighbors) and zero otherwise. Since the network is undirected, $A$
is symmetric \footnote{We also suppose that self-interactions are
not allowed, i.e. $a_{ii}=0$, all $i$.}. For any given node $i$, let
$d_i$ be its degree, i.e. the number of $i$'s neighbors. The extent
to which $i$'s neighborhood is clustered can be measured by the
percentage of pairs of $i$'s neighbors that are themselves
neighbors, i.e. by the ratio between the number of triangles in the
graph $G$ with $i$ as one vertex (labeled as $t_i$) and the number
of all possible triangles that $i$ could have formed (that is,
$T_i=d_i(d_i-1)/2$) \footnote{From now on we will assume that the
denominators of CCs are well-defined. If not, we will simply set the
CC to zero.}. It is easy to see that the CC for node $i$ in this
case reads:

\begin{equation}
C_i(A)=\frac{\frac{1}{2}\sum_{j\neq i}\sum_{h\neq
(i,j)}{a_{ij}a_{ih}a_{jh}}}{\frac{1}{2}d_i(d_i-1)}=\frac{(A^3)_{ii}}{d_i(d_i-1)},
\label{Eq:CC_BUN}
\end{equation}
where $(A^3)_{ii}$ is the $i$-th element of the main diagonal of
$A^3=A\cdot A\cdot A$. Each product $a_{ij}a_{ih}a_{jh}$ is meant to
count whether a triangle exists or not around $i$. Notice that the
order of subscripts is irrelevant, as all entries in $A$ are
symmetric. Of course, $C_i\in[0,1]$. The overall (network-wide) CC
for the graph $G$ is then obtained by averaging $C_i$ over the $N$
nodes, i.e. $C=N^{-1}\sum_{i=1}^{N}{C_i}$. In the case of a random
graph where each link is in place with probability $p\in(0,1)$, one
has that $E[C]=p$ ($E$ stands for the expectation operator).

Binary networks treat all edges present in $G$ as they were
completely homogeneous. More recently, scholars have become
increasingly aware of the fact that real networks exhibit a relevant
heterogeneity in the capacity and intensity of their connections
\cite{Barr04,Barr05,Bart05,DeMontis2005,Kossinets2006,Onnela2007}.
Allowing for this heterogeneity might be crucial to better
understand the architecture of complex networks. In order to
incorporate such previously neglected feature, each edge $ij$
present in $G$ (i.e. such that $a_{ij}=1$) is assigned a value
$w_{ij}>0$ proportional to the weight of that link in the network.
For example, weights can account for the amount of trade volumes
exchanged between countries (as a fraction of their gross domestic
product), the number of passengers travelling between airports, the
traffic between two Internet nodes, the number of e-mails exchanged
between pairs of individuals, etc.. Without loss of generality, we
can suppose that $w_{ij}\in [0,1]$  \footnote{If some $w_{ij}>1$,
one can divide all weights by $max_{i,j}\{w_{ij}\}$.}. A
\textit{weighted} undirected network (WUN) is thus characterized by
its $N\times N$ symmetric weight matrix $W=\{w_{ij}\}$, where
$w_{ii}=0$, all $i$. Many network measures developed for BUNs have a
direct counterpart in WUNs. For example, the concept of node degree
can be replaced by that of node \textit{strength} \cite{Barr04}:

\begin{equation}
s_i=\sum_{j\neq i}{w_{ij}}. \label{Eq:strength}
\end{equation}

For more complicated measures, however, extensions to WUNs are not
straightforward. To generalize the CC of node $i$ to WUNs, one has
indeed to take into account the weight associated to edges in the
neighborhood of $i$. There are many ways to do that
\cite{Saramaki2006}. For example, suppose that a triangle $ihj$ is
in place. One might then consider only weights of the edges $ih$ and
$ij$ \cite{Barr04}. Alternatively, one might employ the weights of
all the edges in the triangle. In turn, the total contribution of a
triangle can be defined as the geometric mean of its weights
\cite{Onnela2005} or simply as the product among them
\cite{Holme2004,Zhang2005,Grindrod2002,Ahnert2006}. In what follows,
we will focus on the extension of the CC to WUNs originally
introduced in \cite{Onnela2005}:

\begin{equation}
\tilde{C}_i(W)=\frac{\frac{1}{2}\sum_{j\neq i}\sum_{h\neq
(i,j)}{w_{ij}^\frac{1}{3}w_{ih}^\frac{1}{3}w_{jh}^\frac{1}{3}}}{\frac{1}{2}d_i(d_i-1)}=
\frac{(W^{\left[\frac{1}{3}\right]})_{ii}^{3}}{d_i(d_i-1)},
\label{Eq:CC_WUN}
\end{equation}
where we define
$W^{\left[\frac{1}{k}\right]}=\{w_{ij}^\frac{1}{k}\}$, i.e. the
matrix obtained from $W$ by taking the $k$-th root of each entry. As
discussed in \cite{Saramaki2006}, the measure $\tilde{C}_i$ ranges
in $[0,1]$ and reduces to $C_i$ when weights become binary.
Furthermore, it takes into account weights of all edges in a
triangle (but does not consider weights not participating in any
triangle) and is invariant to weight permutation for one triangle.
Notice that $\tilde{C}_i=1$ only if the neighborhood of $i$ actually
contains all possible triangles that can be formed and each edge
participating in these triangles has unit (maximum) weight. Again,
one can define the overall clustering coefficient for WUNs as
$\tilde{C}=N^{-1}\sum_{i=1}^{N}{\tilde{C}_i}$.

In this paper we discuss extensions of the CC for BUNs and WUNs
(eqs. \ref{Eq:CC_BUN} and \ref{Eq:CC_WUN}) to the case of directed
networks. It is well-known that many real-world complex networks
involve non-mutual relationships, which imply non-symmetric
adjacency or weight matrices. For instance, trade volumes between
countries \cite{Garla2004,Garla2005,Serrano2003} are implicitly
directional relations, as the export from country $i$ to country
$j$ is typically different from the export from country $j$ to
country $i$ (i.e. imports of $i$ from $j$). If such networks are
symmetrized (e.g., by averaging imports and exports of country
$i$), one could possibly underestimate important aspects of their
network architecture.

\begin{figure}[h]
        \centering {\includegraphics[width=8cm]{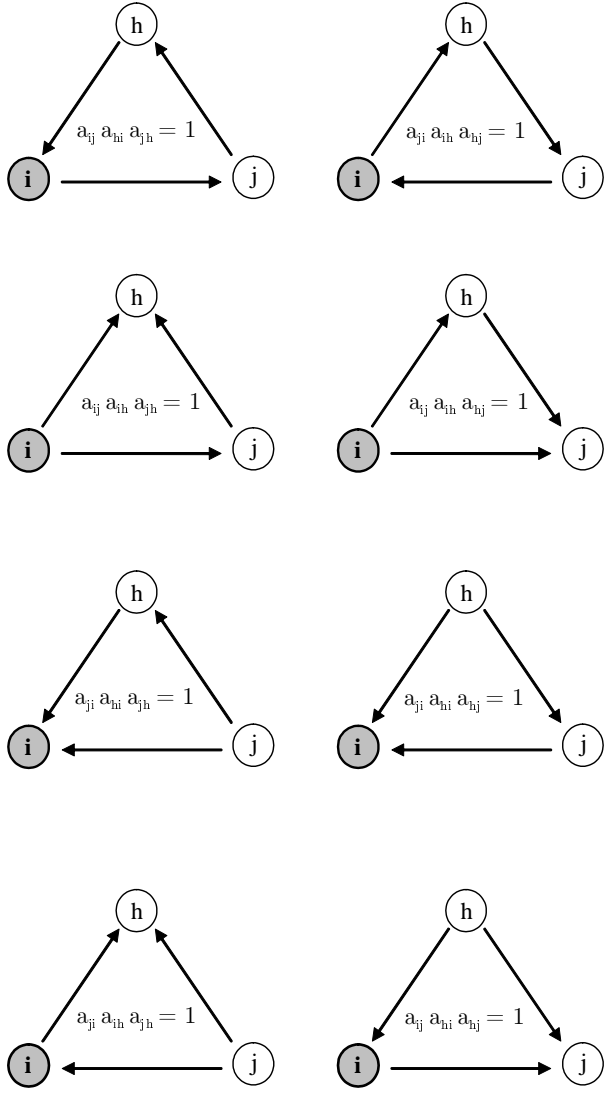}}
        \caption{Binary directed graphs. All 8 different triangles with
        node $i$ as one vertex. Within each triangle is reported the product
        of the form $a_{\ast \ast}a_{\ast \ast}a_{\ast \ast}$ that works as
        indicator of that triangle in the network.}\label{Fig:triangles}
        \end{figure}

Alternative extensions of the CC to weighted or directed networks
have been recently introduced in the literature on ``network
motifs'' \footnote{That is, sets of topologically-equivalent
subgraphs of a network.}. As mentioned, \cite{Onnela2005}
generalizes the CC to weighted -- and possibly directed -- networks.
Similarly, \cite{Milo2002} compute the recurrence of all types of
three-node connected subgraphs in a variety of real-world
\textit{binary} directed networks from biochemistry, neurobiology,
ecology and engineering. However, the weighted CC in
\cite{Onnela2005} does not explicitly discriminate between different
directed triangles (cf. Figure \ref{Fig:triangles}), while
\cite{Milo2002} do not allow for a weighted analysis. This work
attempts to bridge the two latter approaches and presents a unifying
framework where, in addition to the measures already discussed in
\cite{Onnela2005,Milo2002}, one is able to: (i) explicitly account
for directed \textit{and} weighted links; and (ii) define a
weighted, directed version of the CC for any type of triangle
pattern (i.e., three-node connected subgraph). To compute such
coefficients, we shall employ the actual and potential number of
directed-triangle patterns of any given type.

\bigskip

\noindent \textit{Preliminaries}. In directed networks, edges are
oriented and neighboring relations are not necessarily symmetric. In
the case of binary directed networks (BDNs), we define the
\textit{in-degree} of node $i$ as the number of edges pointing
towards $i$ (i.e., inward edges). The \textit{out-degree} of node
$i$ is accordingly defined as the number of edges originating from
$i$ (i.e., outward edges). Formally:

\begin{eqnarray}
d_i^{in}=\sum_{j\neq i}{a_{ji}}=(A^T)_i \textbf{1}\\
d_i^{out}=\sum_{j\neq i}{a_{ij}}=(A)_i \textbf{1},
\end{eqnarray}
where $A^T$ is the transpose of $A$, $(A)_i$ stands for the $i$-th
row of $A$, and $\textbf{1}$ is the $N$-dimensional column vector
$(1,1,\dots,1)^T$. The \textit{total-degree} of a node i simply
the sum of its in- and out-degree:

\begin{equation}
d_i^{tot}=d_i^{in}+d_i^{out}=(A^T+A)_i\textbf{1}.
\end{equation}
Finally, provided that no self interactions are present, the number
of bilateral edges between $i$ and its neighbors (i.e. the number of
nodes $j$ for which both an edge $i\rightarrow j$ and an edge
$j\rightarrow i$ exist) is computed as:

\begin{equation}
d_i^{\leftrightarrow}=\sum_{j\neq i}{a_{ij}a_{ji}}=A^2_{ii}.
\end{equation}
It is easy to see that in BUNs one has:
$d_i=d_i^{tot}-d_i^{\leftrightarrow}$.

The above measures can be easily extended to weighted directed
networks (WDNs), by considering in-, out- and total-strength (see
eq. \ref{Eq:strength}).

\bigskip

\noindent \textit{Binary Directed Networks}. We begin by introducing
the most general extension of the CC to BDNs, which considers all
possible directed triangles formed by each node, no matter the
directions of their edges. Consider node $i$. When edges are
directed, $i$ can generate up to 8 different triangles with any pair
of neighbors \footnote{Of course, by a symmetry argument, they
actually reduce to 4 different distinct patterns (e.g. those in the
first column). We will keep the classification in 8 types for the
sake of exposition.}. Any product of the form $a_{ij}a_{ih}a_{jh}$
captures one particular triangle, see Fig. \ref{Fig:triangles} for
an illustration.

The CC for node $i$ ($C_i^D$) in BDNs can be thus defined (like in
BUNs) as the ratio between all directed triangles actually formed
by $i$ ($t_i^D$) and the number of all possible triangles that $i$
could form ($T_i^D$). Therefore:

\begin{equation*}
C_i^D(A) = \frac{t_i^D}{T_i^D} =
\end{equation*}
\begin{equation}
=\frac{\frac{1}{2}\sum_{j}\sum_{h}{(a_{ij}+a_{ji})
(a_{ih}+a_{hi})(a_{jh}+a_{hj})}}{[d_i^{tot}(d_i^{tot}-1)-2d_i^{\leftrightarrow}]}=
\label{Eq:CC_BDN}
\end{equation}
\begin{equation*}
=\frac{(A+A^T)^3_{ii}}{2[d_i^{tot}(d_i^{tot}-1)-2d_i^{\leftrightarrow}]},
\end{equation*}
where (also in what follows) sums span over $j\neq i$ and $h\neq
(i,j)$. In the first line of eq. (\ref{Eq:CC_BDN}), the numerator
of the fraction is equal to $t_i^D$, as it simply counts all
possible products of the form $a_{ij}a_{ih}a_{jh}$ (cf. Fig.
\ref{Fig:triangles}). To see that
$T_i^D=d_i^{tot}(d_i^{tot}-1)-2d_i^{\leftrightarrow}$, notice that
$i$ can be possibly linked to a maximum of
$d_i^{tot}(d_i^{tot}-1)/2$ pairs of neighbors and with each pair
can form up to 2 triangles (as the edge between them can be
oriented in two ways). This leaves us with
$d_i^{tot}(d_i^{tot}-1)$ triangles. However, this number also
counts ``false'' triangles formed by $i$ and by a pair of directed
edges pointing to the same node, e.g. $i\rightarrow j$ and
$j\rightarrow i$. There are $d_i^{\leftrightarrow}$ of such
occurrences for node $i$, and for each of them we have wrongly
counted two ``false'' triangles. Therefore, by subtracting
$2d_i^{\leftrightarrow}$ from the above number we get $T_i^D$.
This implies that $C_i^D\in [0,1]$. The overall CC for BDNs is
defined as $C^D=N^{-1}\sum_{i=1}^{N}{C_i^D}$.

The CC in eq. (\ref{Eq:CC_BDN}) has two nice properties. First, if
$A$ is symmetric, then $C_i^D(A)=C_i(A)$, i.e. it reduces to
(\ref{Eq:CC_BUN}) when networks are undirected. To see this, note
that if $A$ is symmetric then $d_i^{tot}=2d_i$ and
$d_i^{\leftrightarrow}=d_i$. Hence:

\begin{eqnarray}
C_i^D(A) & = & \frac{(2A)^3_{ii}}{2[2d_i(2d_i-1)-2d_i]}= \nonumber \\
         & = & \frac{(A)^3_{ii}}{d_i(d_i-1)}=C_i(A) \label{Eq:CC_BDN_to_BUN}
\end{eqnarray}

Second, the expected value of $C_i^D$ in random graphs, where each
edge is independently in place with probability $p\in(0,1)$ (i.e.
$a_{ij}$ are i.i.d. Bernoulli($p$) random variables), is still $p$
(as happens for BUNs). Indeed, the expected value of $t_i^D$ is
simply $4(N-1)(N-2)p^3$. Furthermore, note that $d_i^{in}\sim
d_i^{out}\sim BIN(N-1,p)$ and $d_i^{tot}\sim BIN(2(N-1),p)$. Hence
$E[d_i^{tot}(d_i^{tot}-1)]=E[d_i^{tot}]^2-E[d_i^{tot}]=2(N-1)(2N-3)p^2$.
Similarly, $E[d_i^{\leftrightarrow}]=(N-1)p^2$, which implies that
$E[T_i^D]=4(N-1)(N-2)p^2$ and finally that $E[C_i^D]=p$.

\bigskip

\noindent \textit{Weighted Directed Networks}. The CC for BDNs
defined above can be easily extended to weighted graphs by
replacing the number of directed triangles actually formed by $i$
($t_i^D$) with its weighted counterpart $\tilde{t}_i^{D}$. Given
eq. (\ref{Eq:CC_WUN}), $\tilde{t}_i^{D}$ can be thus computed by
substituting $A$ with $W^{\left[\frac{1}{3}\right]}$. Hence:

\begin{equation}
\tilde{C}_i^D(W) =\frac{\tilde{t}_i^{D}}{T_i^D} =
\frac{[W^{\left[\frac{1}{3}\right]}+(W^T)^{\left[\frac{1}{3}\right]}]^3_{ii}}{2[d_i^{tot}(d_i^{tot}-1)-2d_i^{\leftrightarrow}]},
\label{Eq:CC_WDN}
\end{equation}
Note that when the graph is binary ($W=A$), then
$(W)^{\left[\frac{1}{3}\right]}=W=A$. Hence,
$\tilde{C}_i^D(A)=C_i^D(A)$. Moreover, if $W$ is a symmetric weight
matrix, then the numerator of $\tilde{C}_i^D(W)$ becomes
$[2W^{\left[\frac{1}{3}\right]}]^3$. By combining this result with
the denominator in eq. (\ref{Eq:CC_BDN_to_BUN}), one has that
$\tilde{C}_i^D(W)=\tilde{C}_i(W)$ for any symmetric $W$
\footnote{The CC in \eqref{Eq:CC_WDN} is similar to that presented
by \cite{Onnela2005,Saramaki2006} but takes explicitly into account
edge directionality in computing the maximum number of directed
triangles ($T_i^D$). Conversely, \cite{Onnela2005,Saramaki2006} set
$T_i^D=d_i(d_i-1)$.}.

To compute expected values of $\tilde{C}_i^D$ in random graphs,
suppose that weights are drawn using the following two-step
algorithm. First, assume that any directed edge $i\rightarrow  j$ is
in place with probability $p$ (independently across all possible
directed edges). Second, let the weight $w_{ij}$ of any existing
directed edge (i.e., in place after the first step) be drawn from an
independent random variable uniformly distributed over (0,1]
\footnote{That is, $w_{ij}$ is a random variable equal to zero with
probability $1-p$ and equal to a $U(0,1]$ with probability $p$. Of
course this admittedly na\"{i}ve assumption is made for mathematical
convenience to benchmark our results in a setup where one is
completely ignorant about the true (observed) weight distribution.
In empirical applications, one would hardly expect observed weights
to follow such a trivial distribution and more realistic assumptions
should be made. For example, expected value of CCs might be computed
by bootstrapping (i.e., reshuffling) the observed empirical
distribution of weights in $W$ across the same topological graph
structure, as defined by the observed adjacency matrix $A$. See also
below.}. In this case, one has that
$E[w_{ij}]^\frac{1}{3}=\frac{3p}{4}$. It easily follows that for
this class of random weighted graphs:

\begin{equation}
E[\tilde{C}_i^D] = E[\tilde{C}_i] =
{\left(\frac{3}{4}\right)^3}p<p.
\end{equation}
The overall CC for WDN is again defined as
$\tilde{C}^D=N^{-1}\sum_{i=1}^{N}{\tilde{C}_i^D}$.

\bigskip

\noindent \textit{Clustering and Patterns of Directed Triangles.}
The CCs for BDNs and WDNs defined above treat all possible
directed triangles as they were the same, i.e. if directions were
irrelevant. In other words, both $C^D$ and $\tilde{C}^D$ operate a
symmetrization of the underlying directed graph in such a way that
the original asymmetric adjacency (resp. weight) matrix $A$ (resp.
$W$) is replaced by the symmetric matrix $A+A^T$ (resp.
$W^{\left[\frac{1}{3}\right]}+(W^T)^{\left[\frac{1}{3}\right]}$).
This means that in the transformed graph, all directed edges are
now bilateral. Furthermore, in binary (respectively, weighted)
graphs, edges that were already bilateral count as two
(respectively, receive a weight equal to the sum of the weights of
the two directed edges raised to 1/3).

However, in directed graphs triangles with edges pointing in
different directions have a completely different interpretation in
terms of the resulting flow pattern. Put it differently, they
account for different network motifs. Looking again at Figure
\ref{Fig:triangles}, it is possible to single out four patterns of
directed triangles from $i$'s perspective. These are: (i)
\textit{cycle}, when there exists a cyclical relation among $i$ and
any two of its neighbors ($i\rightarrow j \rightarrow h \rightarrow
i$, or viceversa); (ii) \textit{middleman}, when one of $i$'s
neighbors (say $j$) both holds an outward edge to a third neighbor
(say $h$) and uses $i$ as a medium to reach $h$ in two steps
\footnote{These patterns can be also labeled as ``broken'' cycles,
where the two neighbors whom $i$ attempts to build a cycle with,
actually invert the direction of the flow.}; (iii) \textit{in},
where $i$ holds two inward edges; and (iv) \textit{out}, where $i$
holds two outward edges.

When one is interested in measuring clustering in directed networks,
it is important to separately account for each of the above
patterns. This can be done by building a CC for each pattern (in
both BDNs and WDNs). As usual, each CC is defined as the ratio
between the number of triangles of that pattern actually formed by
$i$ and the total number of triangles \textit{of that pattern} that
$i$ can possibly form. Each CC will then convey information about
clustering of each different pattern within tightly connected
directed neighborhoods. In order to do that, we recall that the
maximum number of all possible directed triangles that $i$ can form
(irrespective of their pattern) can be decomposed as:

\begin{eqnarray}
T_i^D & = & d_i^{tot}(d_i^{tot}-1)-2d_i^{\leftrightarrow} = \nonumber \\
      & = & [d_i^{in}d_i^{out}-d_i^{\leftrightarrow}] + \nonumber \\
      & + & [d_i^{in}d_i^{out}-d_i^{\leftrightarrow}] +\\
      & + & [d_i^{in}(d_i^{in}-1)] + \nonumber \\
      & + & [d_i^{out}(d_i^{out}-1)]= \nonumber \\
      & = & T_1^D+T_2^D+T_3^D+T_4^D. \nonumber
\end{eqnarray}
Let $\{T_i^{cyc},T_i^{mid},T_i^{in},T_i^{out}\}$ the maximum number
of cycles, middlemen, ins and out that $i$ can form. Inspection
suggests that: $T_i^{cyc}=T_1^D$, $T_i^{mid}=T_2^D$,
$T_i^{in}=T_3^D$ and $T_i^{out}=T_4^D$. To see why, consider for
example $T_i^{cyc}$. In that pattern type (see Figure
\ref{Fig:triangles}, top panels), node $i$ is characterized by one
inward link and one outward link. The maximum number of such
patterns is given by $d_i^{in}d_i^{out}$. Again, this also counts
``false'' triangles, formed by $i$ and by a pair of directed edges
pointing to and from a same node $j$. Therefore, one has to subtract
$d_i^{\leftrightarrow}$ to get $T_i^{cyc}$. Incidentally, notice
that $T_i^{cyc}=T_i^{mid}$. The reason why this is indeed the case
becomes evident when one compares the top and the bottom pairs of
triangle patterns in Figure \ref{Fig:triangles}. Indeed, cycles and
middlemen only differ from the orientation of the link connecting
the partners of the reference node ($i$), which does not affect the
maximum number of triangles that $i$ can form.

In order to count all actual triangles formed by $i$, we notice
that:

\begin{eqnarray}
t_i^D & = & (A+A^T)_{ii} = \nonumber \\
      & = & (A^3)_{ii} + (A A^T A)_{ii} + \\
      & + & (A^TA^2)_{ii} + (A^2A^T)_{ii} = \nonumber\\
      & = & t_1^D+t_2^D+t_3^D+t_4^D. \nonumber
\end{eqnarray}

By letting $\{t_i^{cyc},t_i^{mid},t_i^{in},t_i^{out}\}$ the actual
number of \textit{cycles}, \textit{middlemen}, \textit{ins} and
\textit{outs} formed by $i$, simple algebra reveals that
$t_i^{cyc}=t_1^D$, $t_i^{mid}=t_2^D$, $t_i^{in}=t_3^D$ and
$t_i^{out}=t_4^D$. For example:

\begin{eqnarray}
t_i^{cyc} & = & \frac{1}{2}\sum_{j}\sum_{h}{[a_{ij}a_{jh}a_{hi}+a_{ih}a_{hj}a_{ji}]}= \\
          & = & \frac{1}{2}[A_{(i)}AA^{(i)}+A^T_{(i)}A^T(A^T)^{(i)}]= A_{(i)}AA^{(i)}=A^3_{(ii)}. \nonumber
\end{eqnarray}
Similarly:
\begin{eqnarray}
t_i^{mid} & = & \frac{1}{2}\sum_{j}\sum_{h}{[a_{ij}a_{hj}a_{hi}+a_{ih}a_{jh}a_{ji}]}= \\
          & = & \frac{1}{2}[A^T_{(i)}A (A^T)^{(i)}+A_{(i)}A^T A^{(i)}]= A_{(i)}A^T A^{(i)}=(A A^T A)_{(ii)}. \nonumber
\end{eqnarray}
Notice that although $T_i^{cyc}=T_i^{mid}$, now $t_i^{cyc}\neq
t_i^{mid}$ as long as $A$ is asymmetric.

Summing up, we can define a CC for each pattern as follows:

\begin{equation}
C_i^{\ast}=\frac{t_i^{\ast}}{T_i^{\ast}}, \label{Eq:CC_tri_BDN}
\end{equation}
where $\{\ast\}=\{cyc,mid,in,out\}$.

In the case of weighted networks, it is straightforward to replace
$t_i^{\ast}$ with its weighted counterpart $\tilde{t}_i^{\ast}$,
where the adjacency matrix $A$ has been replaced by
$W^{\left[\frac{1}{3}\right]}$. We then accordingly define:

\begin{equation}
\tilde{C}_i^{\ast}=\frac{\tilde{t}_i^{\ast}}{T_i^{\ast}},
\label{Eq:CC_tri_WDN}
\end{equation}
where $\{\ast\}=\{cyc,mid,in,out\}$. To summarize the above
discussion, we report in Table \ref{Tab:TriPatterns} a taxonomy of
all possible triangles with related measures for BDNs and WDNs.

Two remarks on equations (\ref{Eq:CC_tri_BDN}) and
(\ref{Eq:CC_tri_WDN}) are in order. First, note that, for
$\{\ast\}=\{cyc,mid,in,out\}$: (i) when $A$ is symmetric,
$C_i^{\ast}=C_i$; (ii) when $W$ is binary,
$\tilde{C}_i^{\ast}=C_i^{\ast}$ ; (iii) when $W$ is symmetric,
$\tilde{C}_i^{\ast}=\tilde{C}_i$. Second, in random graphs one still
has that $E[C_i^{\ast}]=p$ and
$E[\tilde{C}_i^{\ast}]=\left(\frac{3}{4}\right)^3 p$.

Finally, network-wide clustering coefficients $C^{\ast}$ and
$\tilde{C}^{\ast}$ can be built for any triangle pattern
$\{cyc,mid,in,out\}$ by averaging individual coefficients over the
$N$ nodes.

These aggregate coefficients can be employed to compare the
relevance of, say, cycle-like clustering among different networks,
but not to assess the relative importance of cycle-like and
middlemen-like clustering within a single network. In order to
perform within-network comparisons, one can instead compute the
fraction of all triangles that belong to the pattern $\{\ast\} \in
\{cyc,mid,in,out\}$ in $i$'s neighborhood, that is:
\begin{equation}
f_i^{\ast}=\frac{t_i^{\ast}}{t_i^{D}},
\tilde{f}_i^{\ast}=\frac{\tilde{t}_i^{\ast}}{\tilde{t}_i^{D}}.
\label{Eq:CC_freq_tri}
\end{equation}
and then averaging them out over all nodes. Since for $\{\ast\} \in
\{cyc,mid,in,out\}$ we have that $\sum_{\ast}{f_i^{\ast}}=1$ and
$\sum_{\ast}{\tilde{f}_i^{\ast}}=1$, the above coefficients can be
used to measure the contribution of each single pattern to the
overall clustering coefficient. Notice that, in the case of BDNs,
$f_i^{\ast}$ coefficients simply recover the recurrence of each
pattern in the network, as computed in \cite{Milo2002}.

\bigskip

\noindent \textit{Empirical Application}. The above concepts can be
meaningfully illustrated in the case of the empirical network
describing world trade among countries (i.e., the ``world trade
network'', WTN in what follows). Source data is provided by
\cite{GledData2002} and records, for any given year, imports and
exports from/to a large sample of countries (all figures are
expressed in current U.S. dollars). Here, for the sake of
exposition, we focus on year 2000 only \footnote{That is, the most
recent year available in the database. This also allows us to keep
our discussion similar to that in \cite{Saramaki2006}.}. We choose
to build an edge between any two countries in the WTN if there is a
non-zero trade between them and we assume that edge directions
follow the flow of commodities. Let $x_{ij}$ be $i$'s exports to
country $j$ and $m_{ji}$ imports of $j$ from $i$. In principle,
$x_{ij}=m_{ji}$. Unfortunately, due to measurement problems, this is
not the case in the database. In order to minimize this problem, we
will focus here on ``adjusted exports'' defined as
$e_{ij}=(x_{ij}+m_{ji})/2$ and we build a directed edge from country
$i$ to country $j$ if and only if country $i$'s adjusted exports to
country $j$ are positive. Thus, the generic entry of the adjacency
matrix $a_{ij}$ is equal to one if and only if $e_{ij}>0$ (and zero
otherwise). Notice that, in general, $e_{ij}\neq e_{ji}$. In order
to weight edges, adjusted exports can be tentatively employed.
However, exporting levels are trivially correlated with the ``size''
of exporting/ importing countries, as measured e.g. by their gross
domestic products (GDPs). To avoid such a problem, we firstly assign
each existing edge a weight equal to $\tilde{w}_{ij}=e_{ij}/GDP_i$,
where $GDP_i$ is country $i$'s GDP expressed in 2000 U.S. dollars.
Secondly, we define the actual weight matrix as:

\begin{equation}
W=\{{w}_{ij}\}=\frac{\tilde{w}_{ij}}{max_{h,k=1}^{N}\{\tilde{w}_{hk}\}},
\label{Eq:WTN_weights}
\end{equation}
to have weights in the unit interval. Each entry ${w}_{ij}$ tells us
the extent to which country $i$ (as a seller) depends on $j$ (as a
buyer). The out-strength of country $i$ (i.e. $i$'s exports-to-GDP
ratio) will then measure how $i$ (as a seller) depends on the rest
of the world (as a buyer). Similarly, in-strengths denote how
dependent is the rest of the world on $i$ (as a buyer)
\footnote{Dividing by $GDP_j$ would of course require a
complementary analysis. Notice also that \cite{Saramaki2006} define
adjusted exports as $e(i,j)=e(j,i)=[x(i,j)+m(j,i)+x(j,i)+m(i,j)]/2$,
thus obtaining an undirected binary/weighted network by
construction.}.

\begin{figure}[h]
        \centering {\includegraphics[width=8cm]{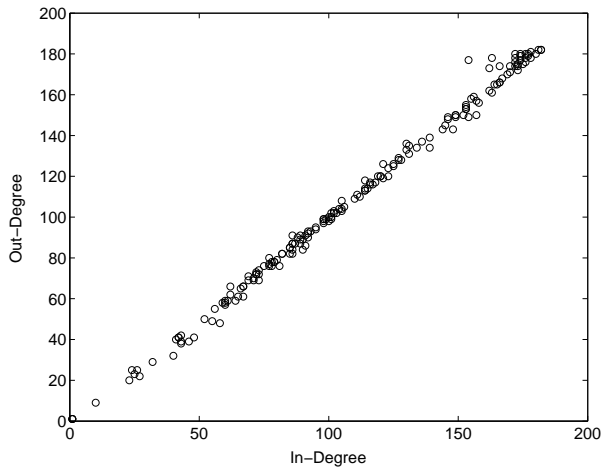}}
        \caption{WTN: In- vs. out-degree in the binary case.}\label{Fig:BDN_InDegree_vs_OutDegree}
        \end{figure}

The resulting WTN comprises $N=187$ nodes / countries and 20105
directed edges. The density is therefore very high
($\delta=0.5780$). As expected, the binary WTN is substantially
symmetric: there is a 0.9978 correlation \footnote{Here and in
what follows, by correlation (or correlation coefficient) between
two variables $X$ and $Y$, we mean the Spearman product-moment
sample correlation, defined as
$\sum_{i}{(x_i-\overline{x})(y_i-\overline{y})}/[(N-1)s_X s_Y]$,
where $s_X$ and $s_Y$ are sample standard deviations. All
correlation coefficients have been computed on original (linear)
data, albeit log-log plots are sometime displayed.} between in-
and out-degree (see Figure \ref{Fig:BDN_InDegree_vs_OutDegree})
and the (non-scaled) $S$-measure introduced in \cite{Fagiolo2006}
is close to zero (0.00397), indicating that the underlying binary
graph is almost undirected.

\begin{figure}[h]
        \centering {\includegraphics[width=8cm]{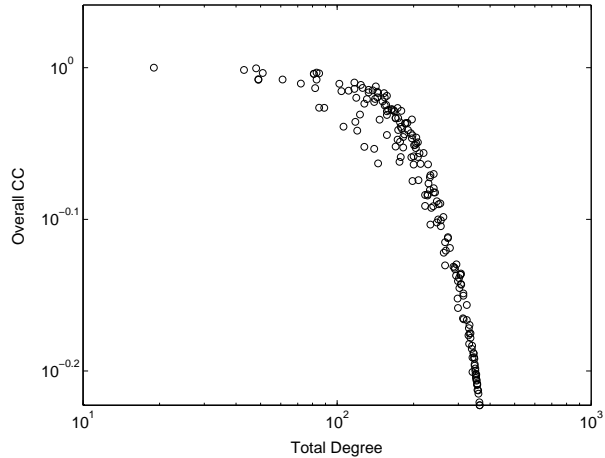}}
        \caption{WTN: Log-log plot of overall directed clustering coefficient vs. total-degree in the binary case.}\label{Fig:BDN_CC_vs_TotDegree}
        \end{figure}

Thus, in the binary case, there seems to be no value added in
performing a directed analysis: since $A$ is almost symmetric, we
should not detect any significant differences among clustering
measures for our four directed triangle patterns. Indeed, we find
that $C^D=0.8125$, while $C^{cyc}=0.8123$, $C^{mid}=0.8127$,
$C^{in}=0.8142$, $C^{out}=0.8108$ \footnote{Accordingly, one has
that $f_i^{cyc}=0.2499$, $f_i^{mid}=0.2501$, $f_i^{in}=0.2531$ and
$f_i^{out}=0.2469$.}. The fact that $C^D>\delta$ also indicates
that the binary (directed) WTN is more clustered than it would be
if it were random (with density $\delta=0.5780$). Finally, Figure
\ref{Fig:BDN_CC_vs_TotDegree} shows that individual CCs ($C_i^D$)
are negatively correlated with total degree ($d_i^{tot}$), the
correlation coefficient being -0.4102. This implies that countries
with few (respectively, many) partners tend to form very
(respectively, poorly) connected clusters of trade relationships.

\begin{figure}[h]
        \centering {\includegraphics[width=8cm]{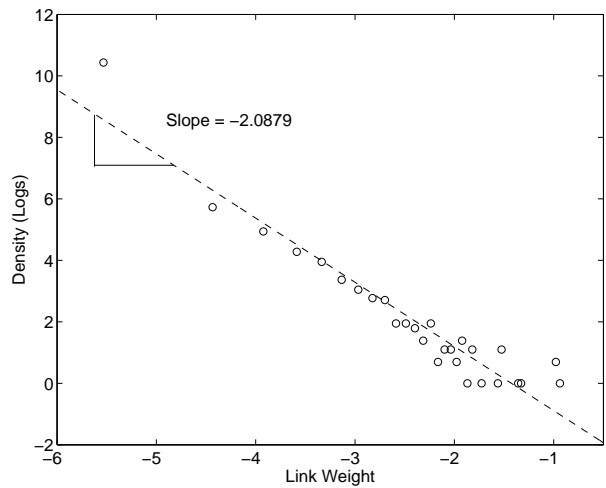}}
        \caption{WTN: Log-log plot of the weight distribution.}\label{Fig:WDN_Weight_Distribution}
        \end{figure}

The binary network does not take into account the heterogeneity of
export flows carried by edges in the WTN. Indeed, when one
performs a WDN analysis on the WTN, the picture changes
completely. To begin with, note that weights ${w}_{ij}$ are on
average very weak (0.0009) but quite heterogeneous (weight
standard deviation is 0.0073). In fact, weight distribution is
very skewed and displays a characteristic power-law shape (see
Figure \ref{Fig:WDN_Weight_Distribution}) with a slope around -2.

\begin{figure}[h]
        \centering {\includegraphics[width=8cm]{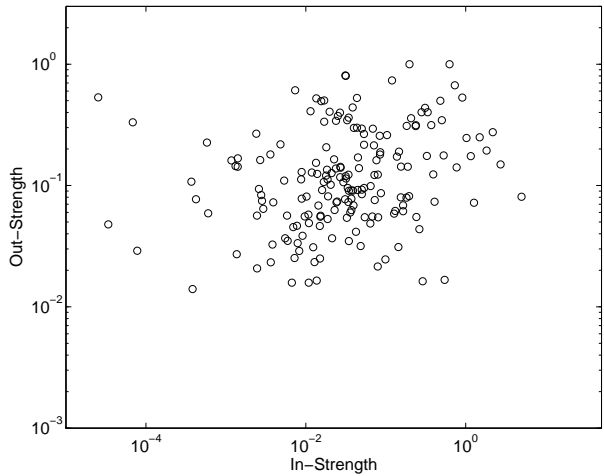}}
        \caption{WTN: Log-log plot of in-strength vs. out-strength.}\label{Fig:WDN_InStr_vs_OutStr}
        \end{figure}

The matrix $W$ is now weakly asymmetric. As Figure
\ref{Fig:WDN_InStr_vs_OutStr} shows, in- and out-strengths are
almost not correlated: the correlation coefficient is 0.09 (not
significantly different from zero). Nevertheless, the (not-scaled)
$S$-measure is still very low (0.1118), suggesting that an
undirected analysis would still be appropriate. We will see now
that, even in this borderline case, a weighted directed analysis of
CCs provides a picture which is much more precise than (and
sometimes at odds with) that emerging in the binary case.

First, unlike in the binary case, the overall average CC
($\tilde{C}^{D}$) is now very low (0.0007) and significantly smaller
than its expected value (0.2438) in random graphs (with the same
density $\delta=0.5780$, but independently, uniformly-distributed
weights). Notice, however, that $\tilde{C}^{D}$ is almost equal to
its expected value in directed graphs characterized by the same
topology (as defined by the adjacency matrix $A$) \textit{but} the
same weight distribution (as defined by the non-zero elements in
$W$), which turns out to be equal to 0.0005 (with a standard
deviation of 0.0001) \footnote{To compute such expected values, we
randomly reshuffled WTN weights in $W$ (by keeping $A$ fixed) and
computed averages/standard deviations of CCs over $M=10000$
independent replications.}.

\begin{figure}[h]
        \centering {\includegraphics[width=8cm]{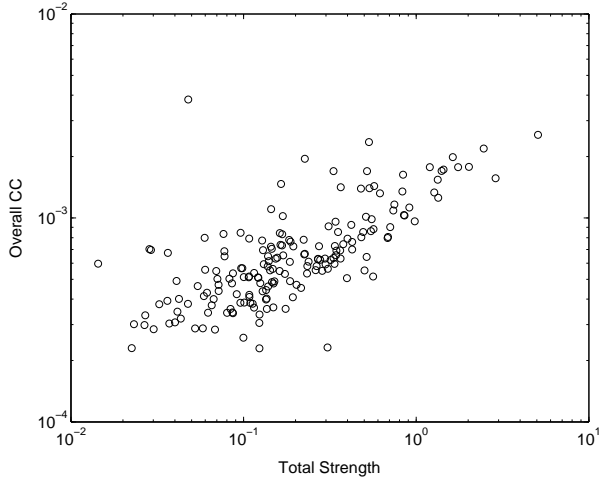}}
        \caption{WTN: Log-log plot of overall CC vs. total strength in the WDN case.}\label{Fig:WDN_CC_vs_TotStr}
        \end{figure}

\begin{figure}[h]
        \centering {\includegraphics[width=8cm]{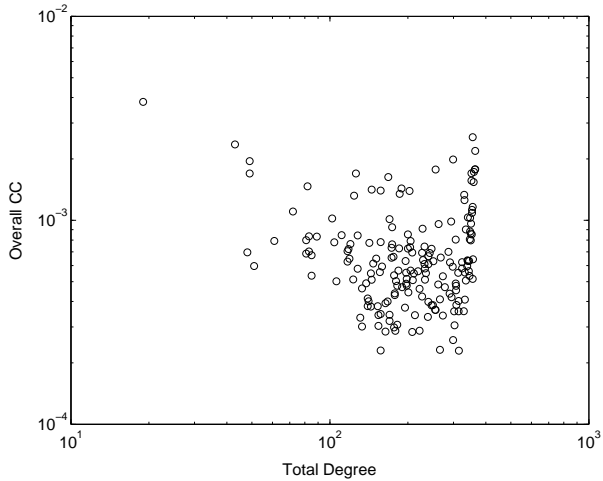}}
        \caption{WTN: Log-log plot of overall CC vs. total degree in the WDN case.}\label{Fig:WDN_CC_vs_TotDegree}
        \end{figure}

Second, $\tilde{C}_i^{D}$ is now positively correlated with total
strength (the correlation coefficient is 0.6421), cf. Figure
\ref{Fig:WDN_CC_vs_TotStr}. This means that, when weight
heterogeneity is taken into account, the implication we have drawn
in the binary case is reversed: countries that are more strongly
connected tend to form more strongly connected trade circles.
Indeed, $\tilde{C}_i^{D}$ exhibits an almost null correlation with
total degree, see Figure \ref{Fig:WDN_CC_vs_TotDegree}.

\begin{figure}[h]
        \centering {\includegraphics[width=8cm]{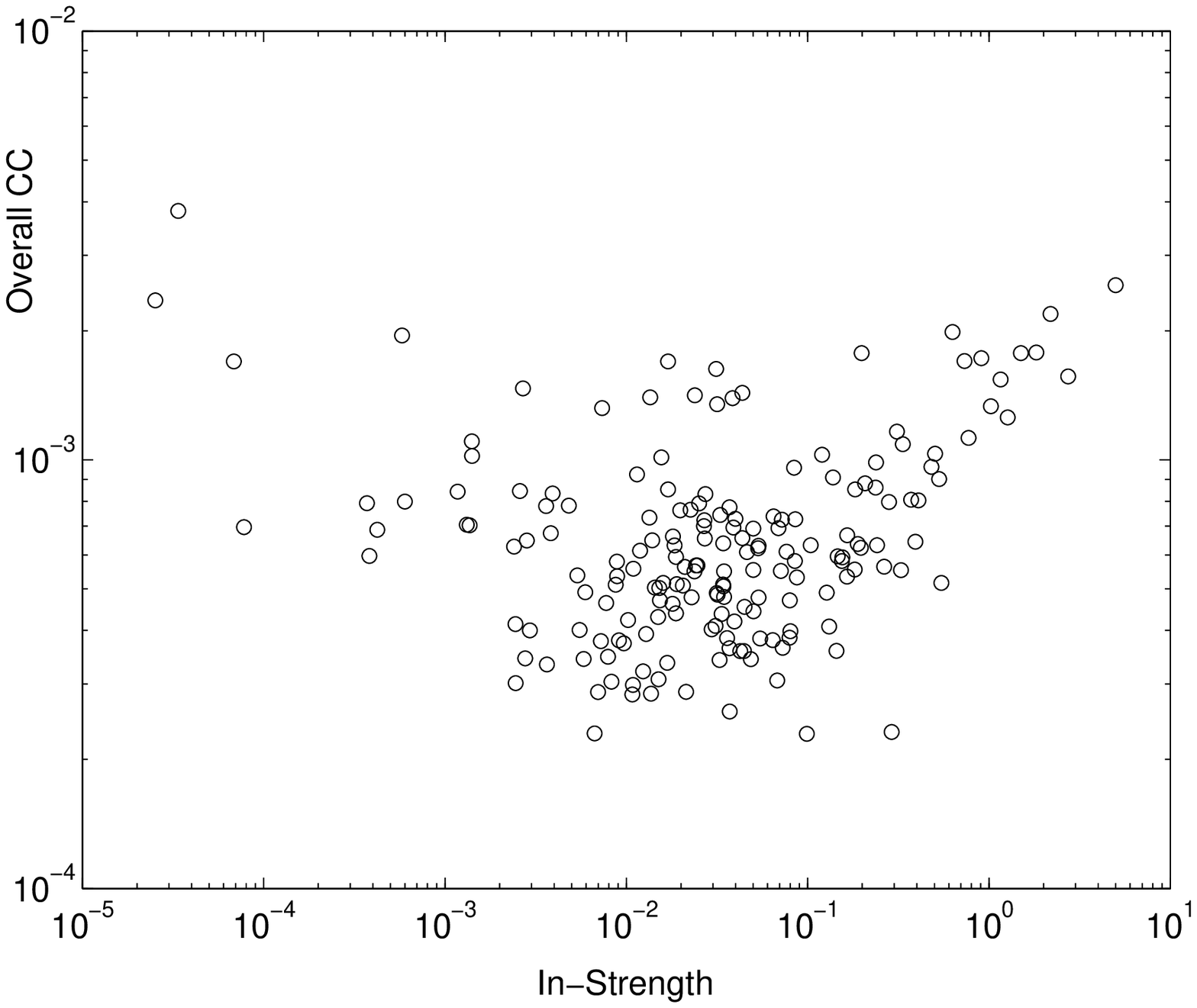}}
        \caption{WTN: Log-Log plot of overall CC vs. in-strength in the WDN case.}\label{Fig:WDN_CC_vs_InStr}
        \end{figure}

Third, despite the weighted network is only weakly asymmetric,
there is a substantial difference in the way clustering is coupled
with exports and imports. $\tilde{C}_i^{D}$ is almost uncorrelated
with in-strength (Figure \ref{Fig:WDN_CC_vs_InStr}), while a
positive slope is still in place when $\tilde{C}_i^{D}$ is plotted
against out-strength. Hence, the low clustering level of weakly
connected countries seems to depend mainly on their weakly
exporting relationships.

Fourth, weighted CC coefficients associated to different triangle
patterns now show a relevant heterogeneity: $\tilde{C}^{\ast}$
range from 0.0004 (cycles) to 0.0013 (out). In addition, cycles
only account for 18\% of all triangles, while the other three
patterns account for about 27\% each. Therefore, countries tend to
form less frequently trade cycles, possibly because they involve
economic redundancies.

\begin{figure}[h]
        \centering {\includegraphics[width=8cm]{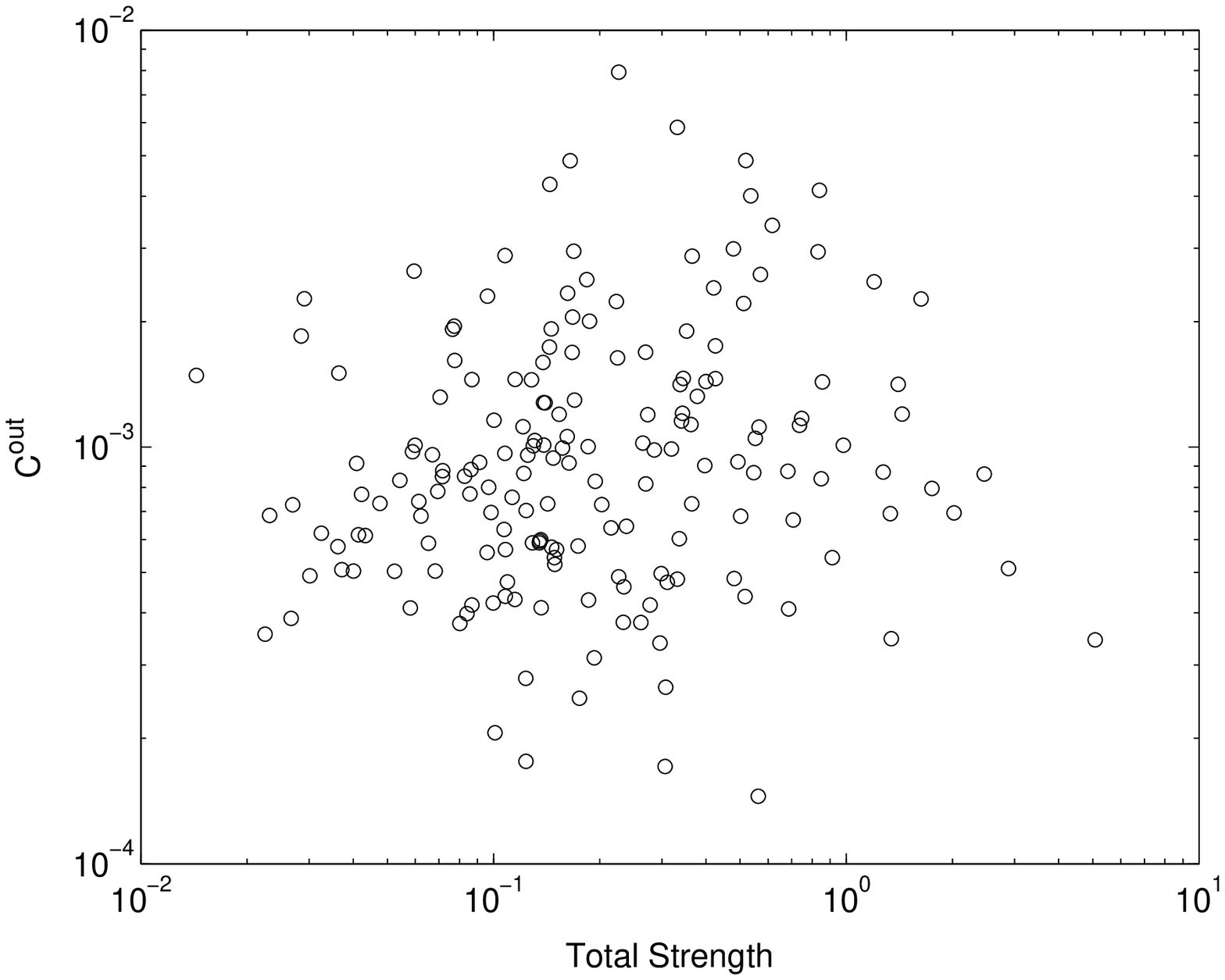}}
        \caption{WTN: Log-log plot of $\tilde{C}_i^{out}$ vs. total strength in the WDN case.}\label{Fig:WDN_CC_out_vs_TotStr}
        \end{figure}

Finally, CCs for different triangle patterns correlate with
strength measures in different ways. While $\tilde{C}_i^{cyc}$,
$\tilde{C}_i^{mid}$ and $\tilde{C}_i^{in}$ are positive and
strongly correlated with total strength, $\tilde{C}_i^{out}$ is
not, see Figure \ref{Fig:WDN_CC_out_vs_TotStr}: countries tend to
maintain exporting relationships with connected pairs of partners
independently of the total strength of their trade circles.

\bigskip

\noindent \textit{Concluding remarks}. In this paper, we have
extended the clustering coefficient (CC), originally proposed for
binary and weighted undirected graphs, to directed networks. We
have introduced different versions of the CC for both binary and
weighted networks. These coefficients count the number of
triangles in the neighborhood of any node independently of their
actual pattern of directed edges. In order to take edge
directionality fully into account, we have defined specific CCs
for each particular directed triangle pattern (cycles, middlemen,
ins and outs). For any CC, we have also provided its expected
value in random graphs. Finally, we have illustrated the use of
directed CCs by employing world trade network (WTN) data. Our
exercises show that directed CCs can describe the main clustering
features of the underlying WTN's architecture much better than
their undirected counterparts.

%\begin{table*}
\begin{sidewaystable}
\caption{A taxonomy of the patterns of directed triangles and their
associated clustering coefficients. For each pattern, we show the
graph associated to it, the expression that counts how many
triangles of that pattern are actually present in the neighborhood
of $i$ ($t_i^\ast$), the maximum number of such triangles that $i$
can form ($T_i^\ast$), for $\ast=\{cyc,mid,in,out,D\}$, and the
associated clustering coefficients for BDNs and WDNs. \textit{Note}.
In the last column:
$\hat{W}=W^{\left[\frac{1}{3}\right]}=\{w_{ij}^{\frac{1}{3}}\}$.}
\begin{tabular}{|c|c|c|c|c|c|}
  \hline
   &  &  &  &  &  \\
  \textbf{Patterns} & \textbf{Graphs} & $t_i^\ast$ & $T_i^\ast$ & \textbf{CCs for BDNs} & \textbf{CCs for WDNs}\\
   &  &  &  &  &  \\
  \hline
  \hline
   &  &  &  &  &  \\
   \raisebox{6ex}[0pt]{Cycle} & \includegraphics[height=2cm]{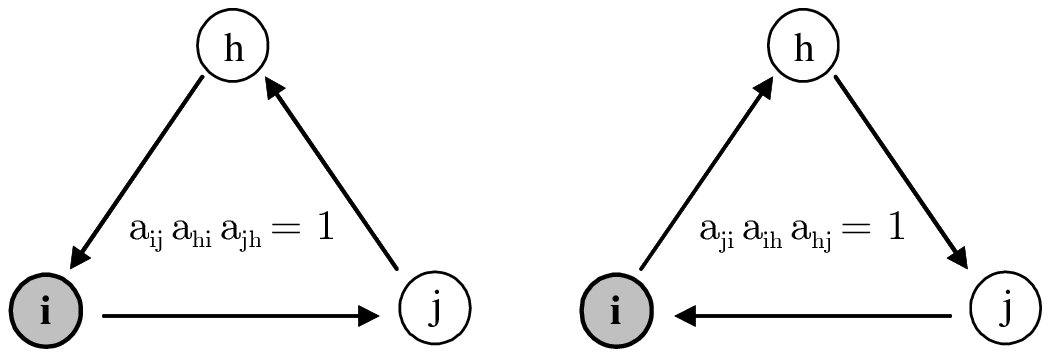} & \raisebox{6ex}[0pt]{$(A)_{ii}^3$} & \raisebox{6ex}[0pt]{$d_i^{in}d_i^{out}-d_i^{\leftrightarrow}$} & \raisebox{6ex}[0pt]{$C_i^{cyc}=\frac{(A)_{ii}^3}{d_i^{in}d_i^{out}-d_i^{\leftrightarrow}}$}  & \raisebox{6ex}[0pt]{$\tilde{C}_i^{cyc}=\frac{(\hat{W})_{ii}^3}{d_i^{in}d_i^{out}-d_i^{\leftrightarrow}}$}\\
   \raisebox{6ex}[0pt]{Middleman} & \includegraphics[height=2cm]{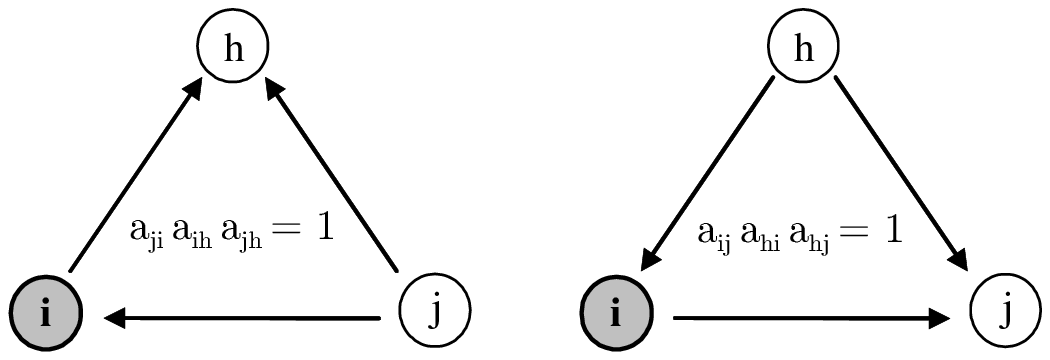} & \raisebox{6ex}[0pt]{$(A A^T A)_{ii}$} & \raisebox{6ex}[0pt]{$d_i^{in}d_i^{out}-d_i^{\leftrightarrow}$} & \raisebox{6ex}[0pt]{$C_i^{mid}=\frac{(A A^T A)_{ii}}{d_i^{in}d_i^{out}-d_i^{\leftrightarrow}}$}  & \raisebox{6ex}[0pt]{$\tilde{C}_i^{mid}=\frac{(\hat{W} \hat{W}^T \hat{W})_{ii}}{d_i^{in}d_i^{out}-d_i^{\leftrightarrow}}$} \\
   \raisebox{6ex}[0pt]{In} & \includegraphics[height=2cm]{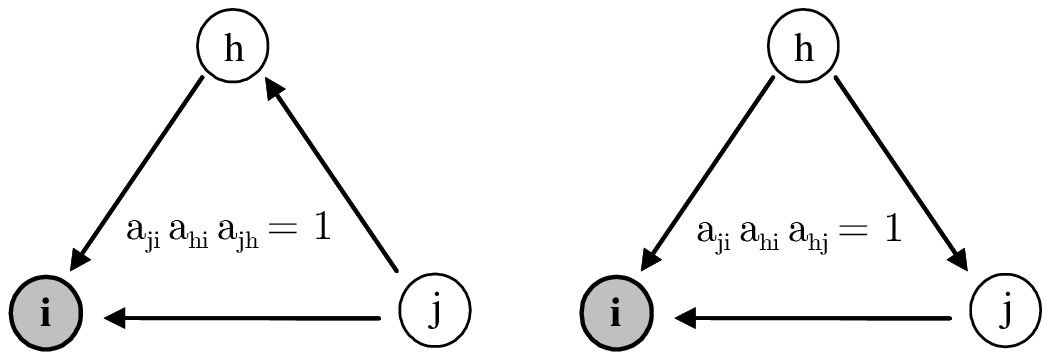} & \raisebox{6ex}[0pt]{$(A^T A^2)_{ii}$} & \raisebox{6ex}[0pt]{$d_i^{in}(d_i^{in}-1)$} & \raisebox{6ex}[0pt]{$C_i^{in}=\frac{(A^T A^2)_{ii}}{d_i^{in}(d_i^{in}-1)}$}  & \raisebox{6ex}[0pt]{$\tilde{C}_i^{in}=\frac{(\hat{W}^T \hat{W}^2)_{ii}}{d_i^{in}(d_i^{in}-1)}$} \\
   \raisebox{6ex}[0pt]{Out} & \includegraphics[height=2cm]{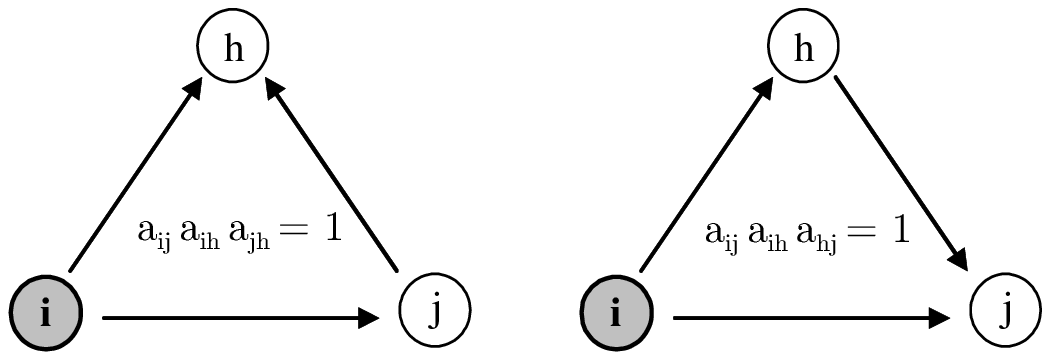} & \raisebox{6ex}[0pt]{$(A^2 A^T)_{ii}$} & \raisebox{6ex}[0pt]{$d_i^{out}(d_i^{out}-1)$} & \raisebox{6ex}[0pt]{$C_i^{out}=\frac{(A^2 A^T)_{ii}}{d_i^{out}(d_i^{out}-1)}$} & \raisebox{6ex}[0pt]{$\tilde{C}_i^{out}=\frac{(\hat{W}^2 \hat{W}^T)_{ii}}{d_i^{out}(d_i^{out}-1)}$} \\
   &  &  &  &  &  \\
    \hline \hline
   &  &  &  &  &  \\
   All (D) & All 8 graphs above & $\frac{(A+A^T)_{ii}^3}{2}$ & \scriptsize $d_i^{tot}(d_i^{tot}-1)-2d_i^{\leftrightarrow}$ & $C_i^{D}=\frac{(A+A^T)_{ii}^3}{2T_i^D}$ & $\tilde{C}_i^{D}=\frac{(\hat{W}+\hat{W}^T)_{ii}^3}{2T_i^D}$ \\
   &  &  &  &  &  \\
  \hline
\end{tabular}
\label{Tab:TriPatterns}
%\end{table*}
\end{sidewaystable}

\begin{acknowledgments}
\noindent Thanks to Javier Reyes, Stefano Schiavo, and an anonymous
referee, for their helpful comments.
\end{acknowledgments}

\bibliographystyle{apsrev}
%\bibliography{clustering}

\end{document}